\documentclass[aps,prl,twocolumn]{revtex4}
\usepackage{ulem}
\usepackage{color}
\usepackage{graphicx}
\usepackage{epsfig}

\usepackage{bm}
\def\bea {\begin{eqnarray}}
\def\eea {\end{eqnarray}}
\def\ra {\rightarrow}

\def \beq{\begin{equation}}
\def \eeq{\end{equation}}
\def \beqa{\begin{eqnarray}}
\def \eeqa{\end{eqnarray}}

\def \la{\langle}
\def \ra{\rangle}
\def \l{\left(}
\def \r{\right)}
\def \raa{R_{AA}(p_T)}
\def \v2{v_2(p_T)}

\def\xp{x_\perp}

\def\6{\partial}

\def\nn{\nonumber}

\usepackage{xspace}

\newcommand{\sNN}{\sqrt{s_{\rm NN}}}

\begin{document}
\title{Directed Flow of Charm Quarks as a Witness of the Initial Strong Magnetic Field in Ultra-Relativistic Heavy Ion Collisions } 
\author{Santosh K. Das$^{a}$, S. Plumari$^{a,b}$, S. Chatterjee$^{c,d}$, J. Alam$^{c}$, F. Scardina$^{a,b}$  and V. Greco$^{a,b}$ }

\affiliation{$^a$ Department of Physics and Astronomy, University of Catania, 
Via S. Sofia 64, 1-95125 Catania, Italy}
\affiliation{$^b$ Laboratori Nazionali del Sud, INFN-LNS, Via S. Sofia 62, I-95123 Catania, Italy}
\affiliation{$^c$ 
Variable Energy Cyclotron Centre-HBNI, 1/AF, Bidhan Nagar, 
Kolkata - 700064}
\affiliation{$^d$School of Physical Sciences, National Institute of Science Education and Research, Jatni, 752050, India}
\date{}
\begin{abstract}
Ultra-relativistic Heavy-Ion Collision (HIC) generates very strong initial magnetic field ($\vec B$) inducing
a vorticity in the reaction plane. The high $\vec{B}$ influences the evolution dynamics that is 
opposed by the large Faraday current due to electric field generated by the time varying $\vec{B}$. 
We show that the resultant effects entail a significantly large directed flow ($v_1$) of charm quarks (CQs) compared to 
light quarks due to a combination of several favorable conditions for CQs, mainly: (i) unlike light quarks 
formation time scale of CQs, $\tau_f \simeq \, 0.1 \rm fm/c$ is comparable to  the time scale when
$\vec B$ attains its maximum value and  (ii) the kinetic relaxation time of CQs  
is similar to  the QGP lifetime, this helps the CQ to retain the initial kick picked up from the electromagnetic field in the 
transverse direction. The effect is also odd under charge exchange allowing to distinguish it from the vorticity 
of the bulk matter due to the initial angular momentum conservation; conjointly thanks to its mass, 
$M_c >>\Lambda_{QCD}$, there should be no mixing with the chiral magnetic dynamics. Hence 
CQs provide very crucial and independent information on the strength of the magnetic field produced in HIC. 

\vspace{2mm}
\noindent {\bf PACS}: 25.75.-q; 24.85.+p; 05.20.Dd; 12.38.Mh

\end{abstract}
\maketitle

The properties of the hot and dense phase of matter referred to as Quark-Gluon Plasma(QGP) 
expected to be produced in nuclear collisions at relativistic energies 
are governed by light quarks and gluons~\cite{Shuryak:2004cy,Science_Muller}. 
The heavy flavors namely, charm and bottom quarks play crucial
roles in probing the QGP~\cite{hfr}.
The special role of CQs  as a probe of the QGP  resides on the fact that their mass, $M$ is 
significantly larger than the  typical temperatures ($T$) achieved in QGP  
and the QCD scale parameter ($\Lambda_{QCD}$) {\it i.e.} $M \gg T, \Lambda_{QCD}$, therefore,
the production of heavy quarks is 
essentially limited to the primordial stage of a heavy-ion collision with a formation time
$\tau_c \sim 1/2\, M_c \simeq 0.08 \, \rm fm/c$. In such a scenario the probability  of
CQs getting annihilated or created during the evolution is much smaller compared
to light quarks and gluons. As a consequence CQs witness the entire space-time evolution of the system and 
can act as an effective probe of the created matter.

The two main experimental observables related to CQs which have been
extensively used as QGP probes are: (i)  the nuclear suppression 
factor, $R_{AA}$ which is the ratio of
the  $p_T$ spectra of heavy flavored hadrons (D and B) produced 
in nucleus + nucleus collisions to those produced in proton + proton collisions 
(appropriately scaled at a given $\sqrt{s_{NN}}$ and (ii)
the elliptic flow, $v_2=\langle cos(2\phi_p)\rangle$,
a measure of the anisotropy in the angular distribution that corresponds to the anisotropic 
emission of particles with respect to the reaction plane. In  the present work we  demonstrate that
the directed flow $v_1=\langle cos(\phi_p)\rangle=\langle p_x/p_T\rangle$ of CQs is a superior probe 
to estimate the magnetic field generated in non-central HICs.  

Several theoretical efforts have been made within the ambit of Fokker Planck 
~\cite{BS,moore,rappv2,rappprl,Das,hiranov2,Gossiaux:2012ea,
alberico,bass,hees,bassnpa,jeon,he1,ali,vc} and relativistic Boltzmann transport approaches
~\cite{gossiauxv2,gre,Uphoff:2012gb,Younus:2013rja,Zhang:2005ni,Molnar:2006ci,fs,HB,Cao:2016gvr}
to calculate $R_{\mathrm AA}$ ~\cite{stare,phenixe,phenixelat,alice} and $v_2$~\cite{phenixelat}. Essentially all the 
models show some difficulties in simultaneously
describing the $R_{AA}(p_T)$ and $v_2(p_T)$ and such a trait is not
only present at RHIC but also appears in a stronger way at LHC energy.
However it has been shown in Ref.\cite{Das:2015ana} that a nearly constant drag or slightly rising with T 
in the range of $\Gamma \sim \, 0.15-0.2 \, \rm fm^{-1}$ is able to simultaneously describe  $\raa$
and $\v2$ at least at RHIC energy, while at LHC energy it still remains 
uncertain also due to the large experimental error bars (see~\cite{rappv2,rappprl,He:2011qa}).

In recent years it has been recognized  that a very strong magnetic field is created
at early times in heavy ion collisions~\cite{magnetic1, magnetic2}. 
The impact of the magnetic field was explored mainly in relation to
the chiral magnetic effect \cite{Kharzeev:2007jp,Fukushima:2008xe}, but also to jet 
energy loss \cite{Tuchin:2010vs}, to $J/\Psi$ elliptic flow \cite{Guo:2015nsa} as well as to thermal photon and
dilepton productions \cite{Tuchin:2012mf}, and very recently to the CQ diffusion
coefficients \cite{Fukushima:2015wck,Finazzo:2016mhm}.
The  estimated values of the initial field strengths, $eB_y \sim 5\,m^2_{\pi}$ and $\sim 50\,m^2_{\pi}$ at RHIC and 
LHC energies respectively, which is several orders of magnitude higher than the values  predicted at
the surface of magnetars. 
Since the CQs are produced at the early stage of HICs, we argue
that their dynamics will be  affected by such a strong magnetic field and they
will be able to retain these effects till its detection as $D$  mesons in experiments.
The $\vec B$ field generated in non-central HICs is dominated by the $\vec y$-component 
which induces a Faraday current in the $xz$ plane.
In particular due to the expansion along the $z$ axis the Lorentz force is directed along the 
negative (positive) $\vec x$ 
direction in the forward (backward) rapidity region for positively (negatively) charged quarks. 
This can be seen as a classical Hall effect that generates a directed transverse flow.  In addition to the Hall effect
the time dependence of $\vec B$ generates a large electric field due to Faraday effect according to 
$\vec{\nabla} \times \vec E= -\partial \vec B/\partial t $. The induced Faraday current opposes the drift 
due to the magnetic field. The combination of the two effects result in a finite $v_1$.

To study quantitatively the global dynamics of CQs we solve
the relativistic Langevin equation in an expanding QGP background.
In this work the initial conditions for solving the Langevin equation for the CQs and the relativistic transport code for
the  background are constrained by the experimental data on the $\raa$ and $v_2$ of
$D$ meson~\cite{Das:2015ana} and the transverse momentum spectra and the
elliptic flow of the bulk, see ~\cite{Plumari:2015cfa,Ruggieri:2013ova,Ruggieri:2013bda,greco_cascade} for more details.
The dynamics of the CQs in the QGP is largely determined by the drag coefficient, $\Gamma$.
We have set a weak $T$ dependence to $\Gamma$ 
in the interval $0.15-0.2 \rm\, fm^{-1}$ which helps in reproducing
the $R_{AA}(p_T)$  and $v_2(p_T)$ for D meson at  RHIC and LHC collisions
reasonably well~\cite{Das:2015ana}.
The initial condition for the bulk in the transverse $r$-space is taken from the Glauber 
model assuming boost invariance along the longitudinal direction. 
The value of the maximum temperature in the center of the fireball at the initial time, $\tau_0=0.2$ fm/c is set as $T_0=580$ MeV.
The initial spatial and  momentum  distributions of the CQs are set respectively by the 
$N_{col}$ and the FONLL scheme of the charm production in proton+proton collisions~\cite{Cacciari:2005rk,Cacciari:2012ny}
at the same $\sNN$. 

Intensive studies have been performed  in the recent years 
to determine the electromagnetic field generated in ultra-relativistic HIC
\cite{Tuchin:2012mf,Deng:2012pc,Zhong:2014sua}.
In the present work we aim to make the first study on the impact of 
the electromagnetic field on heavy quark dynamics under the assumption of
a constant electrical conductivity ($\sigma_{el}$) of the QGP. This will enable us
to - obtain analytic solutions, highlight the core physics and avoid further numerical complications.
We refer to ~\cite{Gursoy:2014aka}  for  the details of the space-time dependent 
solutions of $\vec{E}$ and $\vec{B}$ (see also~\cite{Tuchin:2012mf,Zhong:2014sua}).
With the $z$  and $x$ axes along the beam and impact parameter directions respectively
the $\vec B$  generated in non-central HIC will point along the $y-$ axis on the average, {\it i.e.}  ${\bf B}=B_y \, \hat e_y$ 
while the other components, $ B_x = B_z = 0$.  The 
electric field, $\bf E$ due to Faraday effect will align along the $x$ axis. 

The density $\rho_\pm(\vec \xp)$ of the protons at $\vec{\xp}$ in the transverse plane can be estimated by 
projecting the probability distribution of the protons homogeneously distributed in a sperichal nucleus moving 
either in the $+ z$  or $-z$ .
In a collision with impact parameter $b\neq 0$ the $+$ and $-$ signs indicate the spectators moving
along $+$ and $-$ $z$ directions respectively and the total magnetic field is the sum of $B_y^{+,-}$ generated by 
$Z$ point-like charges as ~\cite{Kharzeev:2007jp}
\bea
\label{totBy}
e  B_{y} &=&  -Z\int_{-\frac{\pi}{2}}^{\frac{\pi}{2}} d\phi' \int_{x_{\rm in}(\phi')}^{x_{\rm out}(\phi')} d\xp' \xp' \rho_-(\xp') \nonumber \\
{}&&\hspace{-0.2in}\times(eB^+_{y}(\tau,\eta,\xp,\phi) + eB^-_{y}(\tau,\eta,\xp,\phi))\, ,
\eea
where $\phi$ is the azimuthal angle, $\tau \equiv \sqrt{t^2-z^2}$ is the proper time, $\eta \equiv arctan(z/t)$ is the
space-time rapidity, and $x_{\rm in}$ and $x_{\rm out}$ are the endpoints of the $\xp'$ integration 
regions that define the crescent-shaped loci where one finds protons either moving $+$ 
or $-$ $z$ directions but not both. These are given by 
\bea
\label{xpm} 
x_{\rm in/out}(\phi') = \mp \frac{b}{2}\cos(\phi') + \sqrt{R^2-\frac{b^2}{4}\sin^2(\phi')}\, ,
\eea
where $R$ is the radius of the nucleus, and $b$ is the impact parameter of the collision.
The main ingredient of Eq.(\ref{totBy}) is the magnetic field $B_y $ at an arbitrary space-time point 
$(t,\vec \xp, \eta)$ generated by a single charge located at $\xp'$ moving in the $+(-)z$ direction with 
velocity $\vec \beta$.  The $B_y^+$ can be written as,
\bea
\label{Bsingle}
 e B^+_y(\tau,\eta,\xp,\phi) &=& \alpha\sinh(Y_b)(x_\perp \cos\phi -  x'_\perp \cos\phi') \nonumber\\
&& \frac{\sigma_{el}\,\frac{|\sinh(Y_b)|}{2} \xi^{\frac12} +1}{\xi^{\frac32}} e^A\, ,
 \eea
 where $\alpha=e^2/(4\pi)$ is the electromagnetic
 coupling, 
 $Y_b\equiv {\rm \arctan}(\beta)$ is the beam rapidity of the $+$ mover,  $A$ and $\xi$ stand for 
 \bea
 \label{ADelta}
\!\!\!\!\!\!\!\!\!\! A \!\!&\equiv&  \!\!\! {\frac{\sigma_{el}}{2}\left(\tau\sinh(Y_b)\sinh(Y_b-\eta)-
 |\sinh(Y_b)| \xi^{\frac12}\right)} \\  
\!\!\!\!\!\!\!\!\!\!  \xi \!\!&\equiv& \!\!\!\tau^2\sinh^2(Y_b-\eta) + \xp^2+\xp^{'2}\nn\\{}&& - 2\xp\xp'\cos(\phi-\phi')\,  .
\eea

We note that the time evolution of the magnetic field is determined by
$\sigma_{el}$ which drives the magnitude of $A$ in Eq.(\ref{Bsingle}).
The following calculations are performed for $\sigma_{el}=0.023\, \rm fm^{-1}$
obtained from lattice QCD calculations 
\cite{Ding:2010ga,Amato:2013naa,Brandt:2012jc} in the temperature range around
$\sim 2\, \rm T_c$. 

In a similar approach the $x-$component of the electric field produced by
the charges moving along $+$ z direction can be obtained as: 
\bea
\label{Esingle}
 e E^+_x(\tau,\eta,\xp,\phi) = \,  e B^+_y(\tau,\eta,\xp,\phi) \coth(Y_b-\eta)\, ,
\eea
which again has to be convoluted with the transverse charge distribution $\rho_\pm(\xp)$ as
for the magnetic field.
The other components of the electromagnetic field averaged over initial conditions will vanish
or become quite small as in the case of the x-component of the electric field $E_z$ in the
region between the two colliding beams where the plasma is formed.
However, it is has been shown in~\cite{Deng:2012pc}  that the large fluctuations in event by event central collisions 
(not of interest here) can generate  other components of the fields with magnitudes comparable but generally
smaller than to $B_y$ and $E_x$. Also the positive (negative) direction of the $B_y$-field here is conventional
and in the experiments an event-by-event analysis has to be done to find a non vanishing flow.
Furthermore, in principle one should also include the electromagnetic field generated by the
participant protons, however, it has been shown in \cite{Gursoy:2014aka} that its magnitude is 
sub dominant especially in the initial stage that plays  the  leading role for the directed flow considered  here.

\begin{figure}[ht]
\begin{center}
\includegraphics[width=18pc,clip=true]{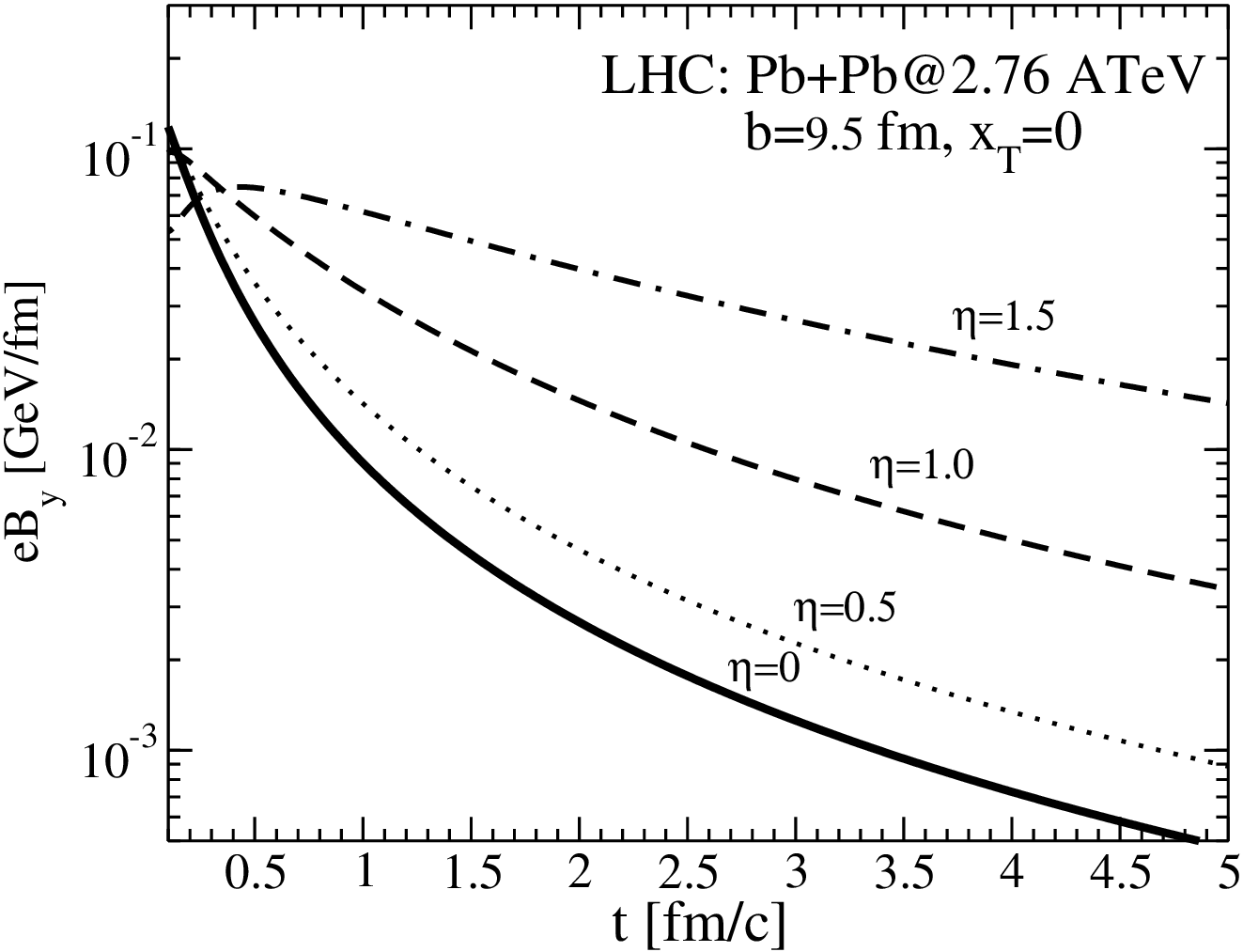}\hspace{2pc}
\includegraphics[width=17.5pc,clip=true]{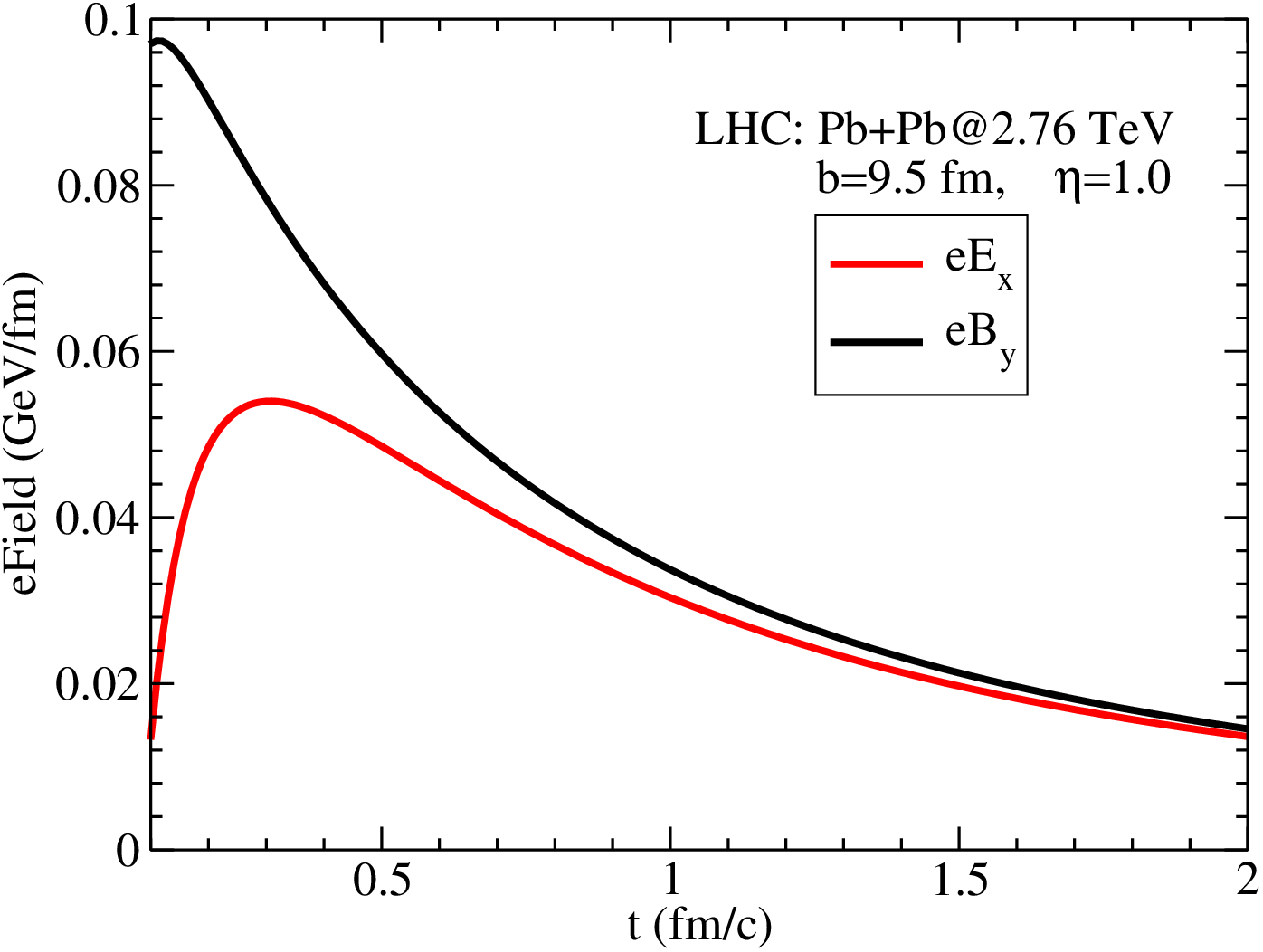}\hspace{2pc}
\caption{(Color online) - Time dependence of  $B_y$ and $E_x$ fields for $\sigma_{el}=0.023 \,\rm fm^{-1}$
in $Pb+Pb$ collisions at ${\sNN}=2.76\,\rm TeV$ for $b=9.5 \, \rm fm$ at $x_\perp=0$.  
Upper panel: $eB_y$ for different space rapidities $\eta$;
Lower panel: time evolution  of both the magnetic field $eB_y$ (black ) and 
the electric field $eE_x$ (red) at forward rapidity $\eta=1.0$.}
\label{fig1}
\end{center}
\end{figure}
In Fig. \ref{fig1} (upper panel) we display the time evolution of the magnetic field 
${\bf B}= B_y \, \hat e_y$ at $\vec \xp= 0$ for various $\eta$
for $Pb+Pb$ at ${\sNN}=2.76 \, \rm TeV$ for  $b=9.5 \, \rm fm$
with $\sigma_{el}=0.023$ fm$^{-1}$ .
The electric field vanishes at this position due to symmetry.
An important factor for sizable directed flow is certainly the fact that
CQs are produced in the early stage. In fact in this region the $B_y$ reduces by about an order
of magnitude  between $t =0.1 \, \rm fm/c$ to  $t=1 \, \rm fm/c$. 
In Fig. \ref{fig1} (lower panel) we depict the time  dependence of both $B_y$ (black line)
and electric field ${\bf E}= E_x \, \hat e_x$ (red line) 
at $\bf \xp=0$ and $\eta=1.0$.  
We note  that for $t< 1 \, \rm fm/c$ there is a large
difference between the $B_y$ and $E_x$, although they become
equal at later time. We will see that this plays an important role in determining the
sign and the size of $v_1$.

The dynamics of the CQs, with charge $q$ and momentum $p$, is governed by the Langevin equation in the 
presence of electromagnetic field, given by
\beqa
\dot{\bf x}\l t\r &=& \frac{\bf p}{E} \\
\dot{\bf p}\l t \r &=& -\Gamma {\bf p}\l t\r + {\bf F}\l t\r + {\bf F}_{ext}\l t\r ,
\label{eq.LE}
\eeqa
where the first term represents the dissipative force and the second term 
represents the fluctuating force $F\l t\r$ regulated by the diffusion coefficient $ D$. 
The third term in Eq.~\ref{eq.LE} represents the external Lorentz force due to the 
electric and magnetic fields. 
We study the evolution with a standard white noise ansatz for $F\l t\r$, {\it i.e}  $\la {\bf F}\l t\r\ra = 0$ and
$\la {\bf F}\l t\r \dot {\bf F}\l t'\r\ra = D\delta\l t-t'\r$.
The ensemble $\la ...\ra$  denotes the averaging of many trajectories for $p$
each consisting of different realizations of $F$ at each time step.  
To  solve the Langevin equation for an expanding system one needs to move to the local rest frame of the
background fluid \cite{hfr,moore}, where an element
moving with velocity $\bf v$ with respect to the laboratory frame will be subjected to  
both $\bf E'$ and $\bf B'$  as determined by Lorentz transformations.
The $F_{ext}$ in the fluid rest frame will be
 \beqa
 {\bf F}_{ext} &=& q{\bf E'} + \frac{q}{E_p}\l {\bf p}\times {\bf B'}\r
 \eeqa
 where $E_p=\sqrt{p^2+M^2}$ is the energy of the heavy quark with momentum $p$.  
    
\begin{figure}[ht]
\begin{center}
\includegraphics[width=19pc,clip=true]{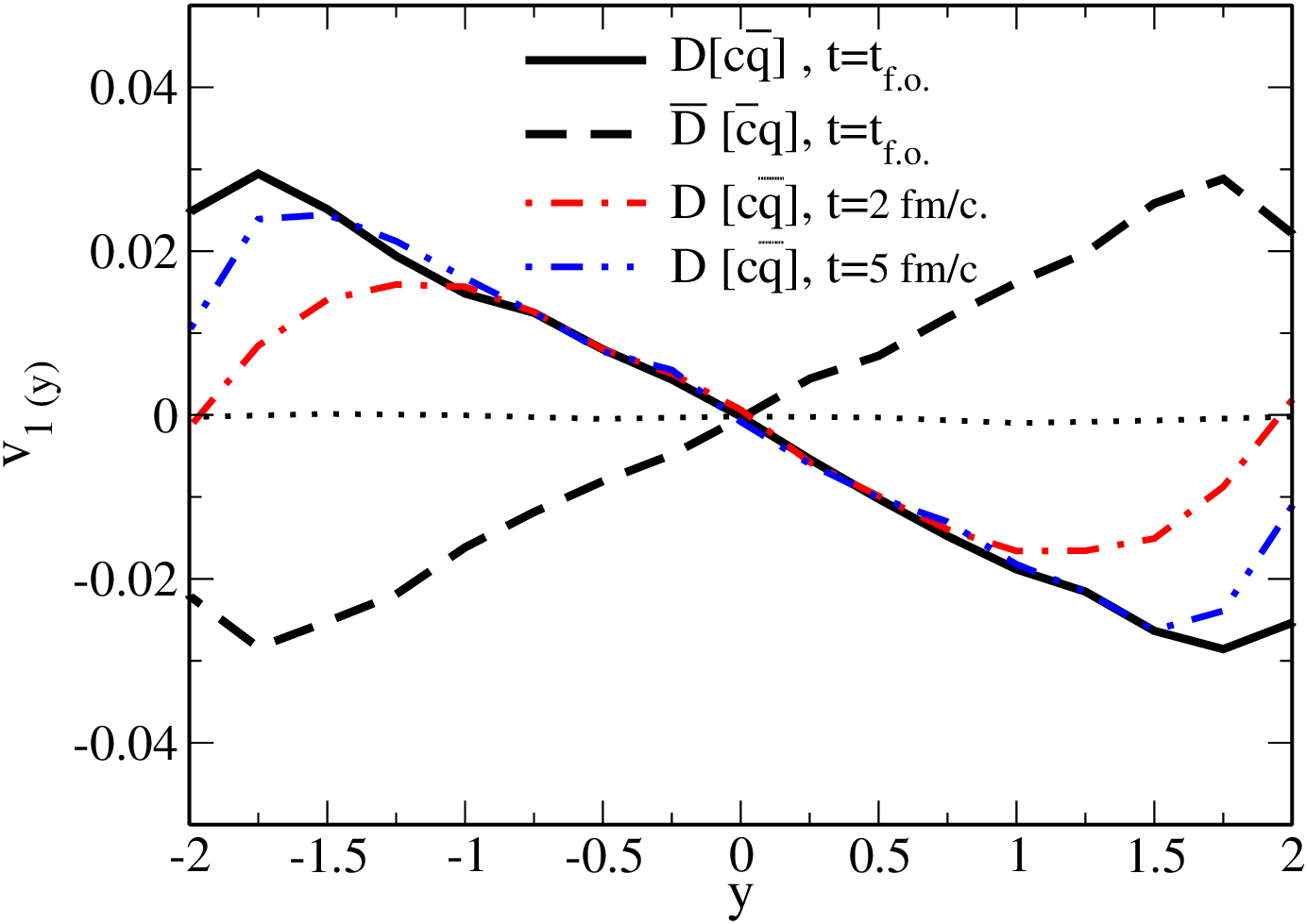}\hspace{2pc}
\caption{(Color line) - Directed flow $v_1$ as a function of the rapidity in $Pb+Pb$ collision at ${\sNN}=2.76 \,\rm TeV$ for 
$b=9.5\, \rm fm$  for D meson [$c\overline q$] at $p_T>1\, \rm GeV$ black (solid) and 
anti-D meson [$\overline c q$] (dashed) line at $t=t_{f.o.}$ (see text). 
Red dash-dot (blue dash-dot-dot) line indicates $v_1$ of $D$ meson at $t=2\,\rm fm/c$ 
($t=5\,\rm fm/c$).}
\label{fig2}
\end{center}
\end{figure}

In Fig. \ref{fig2} we show the resulting directed flow $v_1$ as a function of the rapidity of charm
black (solid line) and anti-charm quarks (dashed line). We can see that there is a substantial $v_1$
at finite rapidity with a peak at $y\simeq \, 1.75$. The flow is negative for positive charged
particle (charm) at forward rapidity which means that the Hall drift induced by the magnetic field
dominates over the displacement caused by the Faraday current associated with the time dependence of
the magnetic field. 
This is a non trivial result and it is partially due to the fact that
the formation time of CQs 
is very close to the time at which the magnetic field attains its maximum, causing a large drift due to Hall effect.

We have followed the dynamics up to $t=12\,\rm fm/c$, but observed that the directed
flow saturates already at $t \simeq 1-2 \, fm/c$ for $|y|<0.5-1$,
and at $t \simeq 5 \, fm/c$ for $|y|< 1.5$ (Fig.\ref{fig2}). Therefore, the slope
$dv_1/dy|_{y=0}\simeq - 1.75\cdot 10^{-2}$ is determined in the very early stage of the collision $t \lesssim 1-2 \, \rm fm/c$.
The time scale of the saturation of $v_1$ as a function of $y$ follows the persistence 
of the $B_y$ and $E_x$ fields with increasing $\eta$ shown in Fig.\ref{fig1} (upper panel).
Therefore, the $v_1$, in particular its slope $dv_1/dy $, is mostly formed in the very early stage 
and its sign and magnitude is essentially controlled by the large value of $B_y$ for $t\lesssim 1.0\,\rm fm/c$.
The predicted value of $v_1(y)$  for D meson [$c\overline q$] is quite large
and the odd behavior 
of $\rm D / \overline D$ doubles the effect that can be measured. Also it would be
a distinctive signal of the electromagnetic field, distinguishable from the
$v_1(y)$ that can be generated by angular momentum conservation as studied in \cite{Becattini:2015ska}. 

It is important to stress that a main feature of the CQs
that turns out to favor the formation of a sizable directed flow is the relative large equilibration
time w.r.t. light quarks. In fact, the relaxation
time of CQs can be estimated as $\tau_{c}^{eq}\simeq \, 1/\Gamma \approx 5-8 \, fm/c$
which is much larger than the light quark and gluon equilibration time, 
$\tau_{QGP}^{eq} \approx \, 0.5-1 \, \rm fm/c$.

\begin{figure}[ht]
\begin{center}
\includegraphics[width=19pc,clip=true]{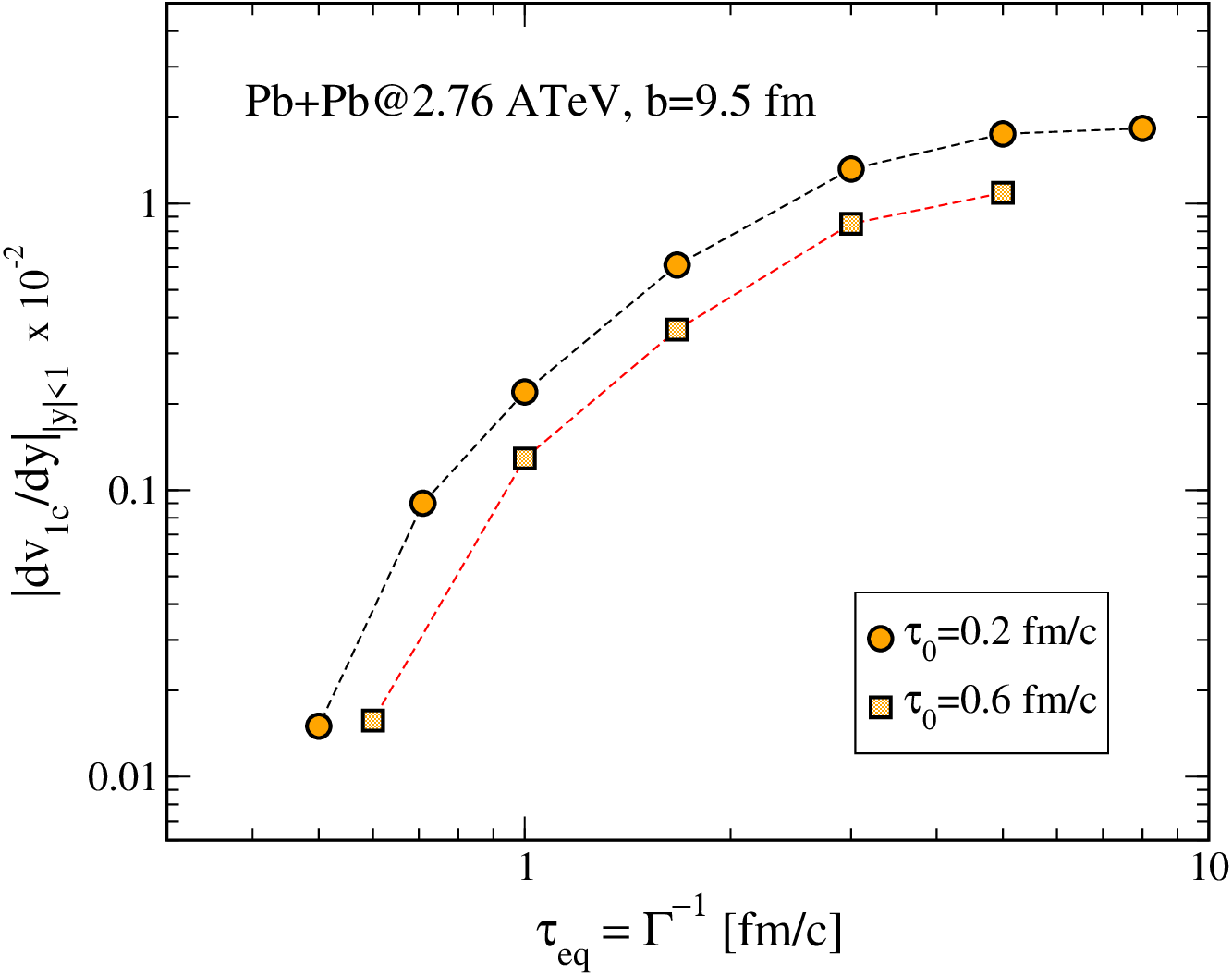}\hspace{2pc}
\caption{(Color online) - Absolute value of the slope of the charm transverse flow $|dv_1/dy|$ around mid-rapidity for
$Pb+Pb$ at ${\sNN}=2.76 \,\rm TeV$ for  $b=9.5\, \rm fm$ as a function of the inverse of the drag coefficient $\Gamma$
and for two different values of the thermalization time  $\tau_0=0.2 \,\rm fm/c$ (circles) and  $\tau_0=0.6 \,\rm fm/c$ (squares).}
\label{fig3-new}
\end{center}
\end{figure}

In Fig. \ref{fig3-new} we show a study of the strong dependence of the transverse flow
on the interaction strength given by the drag coefficient $\Gamma$ and plotted in term
of the equilibration time defined as $\tau_{eq}=1/\Gamma$.

The strong dependence of $v_1$ on $\Gamma$ is evident from the variation of  
$|dv_{1c}/dy|$ with $\tau_{eq}$ as displayed in Fig.~\ref{fig3-new}.
The quantity, $|dv_{1c}/dy|$ for CQ with $\tau_c^{eq} \simeq    
1/\Gamma \approx 5-8 \, fm/c\,$ is at least two orders
of magnitude larger than the corresponding value for light quarks with $\tau_{eq}\sim 0.6$ fm/c 
\cite{Gursoy:2014aka}. This is due to the fact that the transverse kick exerted by the electromagnetic field 
during the time interval, $\tau_{e.m.}$
on the thermalized light quarks (unlike CQ which is out-of-equilibrium) is damped by its random interaction 
in the medium with similar durability. 
However, the lowest points in Fig.\ref{fig3-new} may not be taken as a realistic estimate for $v_1$ of light quarks,
because for that we have to keep in mind at least three other aspects: the dynamics of light quarks cannot
be appropriately studied by using Langevin dynamics as is done usually for heavy quarks, the light hadrons originate abundantly also
from the hadronization of gluons which are not directly affected by the electromagnetic interaction and 
their initial momentum distributions
is quite different from that of CQs. All these aspects cause a further significant 
reduction in the transverse flow of light hadrons which will be discussed in a future work.

In this context it is also important to mention that initially the charm quarks are in a non-equilibrium stage
and due to the scatterings with the medium experience a significant acceleration in the first fm/c.
Such an accelation is proportional to the drag coefficient and it can reach values similar
to the longitudinal expansion rate $1/\tau$ at $\tau_0$ for values of the drag corresponding to
$\tau_{eq}=1/\Gamma \simeq 0.5\, \rm fm/c$. From Fig.\ref{fig3-new} we see that in such a case the strong longitudinal
acceleration implies a nearly complete damping of the transverse kick that would be induced by the magnetic field.

\begin{figure}[ht]
\begin{center}
\includegraphics[width=19pc,clip=true]{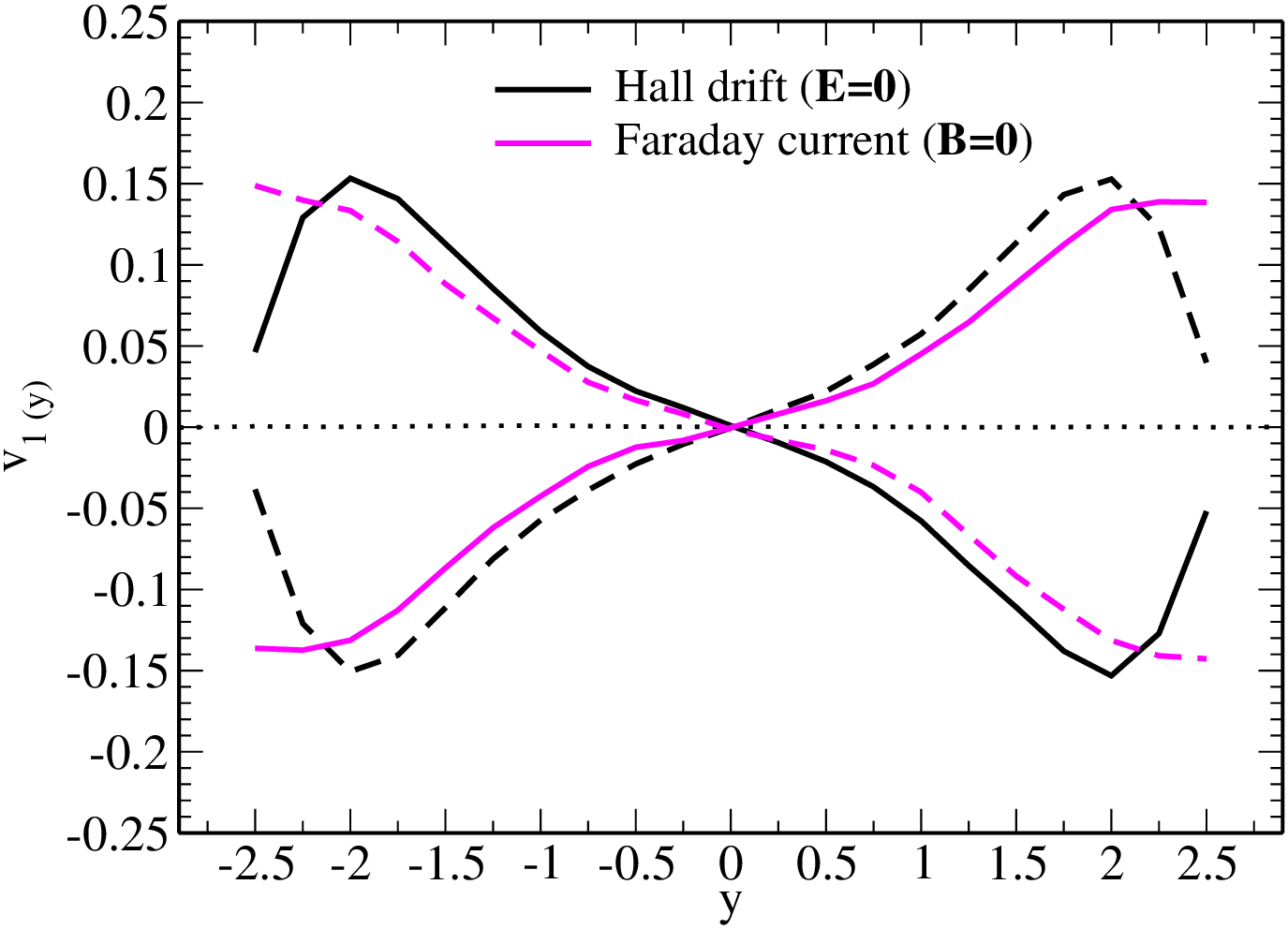}\hspace{2pc}
\caption{(Color online) - Black (Magenta) line shows the variation of  $v_1(y)$ with $y$ generated by the drift due to
Hall effect($B_y \neq 0$, $E_x=0$) (generated by Faraday effect ($ E_x \neq 0$, $B_y=0$)  
in $Pb+Pb$ collision at ${\sNN}=2.76 \,\rm TeV$ for $b=9\, \rm fm $ for $p_T> 1\,\rm GeV$. 
The $v_1(y)$ for charm (anti-charm) is denoted by the solid (dashed) line.}
\label{fig3}
\end{center}
\end{figure}

A last aspect we want to point out is that certainly the strength of the electromagnetic field is important 
to have a sizeable transverse flow, but the underlying dynamics is more subtle.
In Fig. \ref{fig3} we display the $v_1(y)$ that is generated if we switch off artificially
the electric field and keep the action of the magnetic field on (Hall drift only). We notice  that in such a situation  
the $v_1(y)$ (black lines) generated is much larger than the one displayed in Fig.\ref{fig2}.
We observe also that when only the electric field is considered the effect of the Faraday current
generates $v_1$  with opposite sign but a magnitude similar to the Hall drift. 
The $v_1(y)$ in Fig.2 even if not exactly equal to the difference between the Hall drift and the
Faraday current calculated separately, as in Fig. \ref{fig3}, differs from it only by at most $5-10\%$.
We understand that the value of $v_1$ is not 
only decided by the magnitude of the fields, but depends critically on the balance between 
$\vec{E}$ and $\vec{B}$ fields. 
In particular, the magnitude of the magnetic Hall drift, depending on the absolute magnitude
of $B_y$, is large at the formation time of the CQs. This entails a dominance of the Hall drift 
that is kept till the end of the evolution. Instead light quarks likely fail to feel the presence 
of the early high magnitude of $B_y > E_x$ due to their late formation.
In fact, looking at Fig.\ref{fig1} (lower panel) and Fig. \ref{fig3} it is straightforward 
to envisage that if the charged particles
would be produced at $t \simeq 1\, \rm fm/c$ or if the simulation of the dynamics starts at similar
times then the electric and magnetic field nearly compensate their effects, consequently 
$v_1(y)$ with smaller magnitude, see also Fig.\ref{fig3-new}. 

One may also wonder what can be expected for bottom quarks. Granted they have a factor of 2
smaller coupling to the e.m field due to the charge $\pm 1/3$, the larger mass leads to a significant
damping of the Lorentz Force proportional to $p/E_p$. A preliminary calculation shows that
this determines a nearly exact balance between the Hall drift and Faraday current resulting
in a $v_1$ that is about 4-5 times smaller the charm quark one, but its value
critically depends on the details of the drag coefficient, initial time $\tau_0$ and $p_T$ distribution,
that currently under scrutiny and will be presented in a future work.

We have also checked the impact of the electromagnetic field  on $\raa$ and 
$\v2$ and found that the former are not altered by the electromagnetic force;
while an effect of the B-field on $v_2$ can come indirectly from the anisotropy induced in the bulk
as conjectured in \cite{Fukushima:2015wck}.

In summary, the present study suggests that $v_1$ of CQs can be considered as an efficient probe to 
characterize the evolving magnetic field produced in ultra-relativistic HIC. The time evolution
of the field is determined by the electrical conductivity of QGP created in such collisions.  
Our central focus has been to show that the electromagnetic field can generate a sizable $v_1$ for
CQs and hence for D meson,  thanks to several concurring favorable effects for this to happen.
The formation time of CQs is of about $\tau_{form}\sim 0.1\, \rm fm/c$ that is when
the intensity of the $\vec{B}$ field is maximum, even more important aspect is that 
the dynamics at time $t \lesssim 1.0 \rm \,fm/c$ is 
governed by the opposite action of $E_x$ and $B_y$ provides significant amount of net flow. 
Furthermore, the CQs, due to their large relaxation time in contrast  to light quarks,  
are capable of retaining the memory of the initial non-equilibrium dynamics more effectively 
and hence lending stronger signal of the early magnetic dynamics.  
Furthermore, a large number of
light hadrons  originate from the gluons, not directly coupled to
the electromagnetic field. Also in this respect CQ would provide a much cleaner and direct probe
of the magnetic field dynamics. All these favorable conditions largely overwhelm the small suppression
of the Lorentz force by a factor, $p/E_p$ due to their finite mass. In addition, for $M_c >> \Lambda_{QCD}$ 
the CQ dynamics is not significantly affected by the chiral dynamics, therefore, the splitting between 
charge and anti-charge would not mix with the Chiral Magnetic Effect (CME) and/or with possible Chiral 
Vortex Effect (CVE) that can also generate a matter/anti-matter splitting \cite{Jiang:2015cva,Liao:2016diz}.
Thus, CQs would provide an independent way to scrutinize and quantify the initial magnetic field 
which can in turn also contribute to a more quantitative assessment of the CME and CVE.


\vspace{2mm}
\section*{Acknowledgments}
SKD, SP, FS and VG acknowledge the support by the ERC StG under the QGPDyn Grant n. 259684. 
SC thanks XIIth plan project no. 12-R$\&$D-NIS-5.11-0300 and ``Center for Nuclear 
Theory''[PICXII-R$\&$D-VEC-5.02.0500], VECC for support and acknowledges the hospitality received 
during his stay at INFN-LNS when this work was initiated.
VG thanks also the hospitality under the UCAS's PIFI Fellowship project no.2016VMA063 in Beijing, 
where this work was finalized and M. Ruggieri for useful comments on the manuscript.

\end{document}